\documentclass[twocolumn,aps,amssymb,prb,showpacs]{revtex4-1}
\usepackage{graphicx}

\usepackage{amsmath}
\usepackage{float}

\newcommand{\RR}{\right}
\newcommand{\LL}{\left}
\newcommand{\m}{\mathrm}
\newcommand{\xto}{x_0}

 \bigskip
\begin{document}
\title{Macroscopic quantum tunneling in nanoelectromechanical systems}

\author{Mika A. Sillanp\"a\"a, Rapha\"el Khan, Tero T. Heikkil\"a and Pertti J. Hakonen}
\affiliation{Low Temperature Laboratory, Aalto University, P.O. Box 15100, FI-00076 AALTO, Finland}

\begin{abstract}
The experimental observation of quantum phenomena in mechanical degrees of freedom is difficult, as the systems become linear towards low energies and the quantum limit, and thus reside in the correspondence limit. Here we investigate how to access quantum phenomena in flexural nanomechanical systems which are strongly deflected by a voltage. Near a metastable point, one can achieve a significant nonlinearity in the electromechanical potential at the scale of zero point energy. The system can then escape from the metastable state via macroscopic quantum tunneling (MQT). We consider two model systems suspended atop a voltage gate, namely, a graphene sheet, and a carbon nanotube. We find that the experimental demonstration of the phenomenon is currently possible but demanding, since the MQT crossover temperatures fall in the millikelvin range. A carbon nanotube is suggested as the most promising system.
\end{abstract}

\maketitle

\section{Introduction}

The quest towards experimental studies of quantum behavior in nanomechanical systems \cite{LaHaye04,SchwabQ,Kippenberg08} has progressed fast in recent years. A breakthrough took place last year, when the quantum ground state was demonstrated by the Cleland group \cite{ClelandMartinis}, using a 6 GHz piezoelectric mode in resonance with an electrical quantum system made out of Josephson junctions. In general, the difficulty of bringing mechanical degrees of freedom to the quantum limit is due to the challenges posed by several issues. As usual, for quantum-mechanical phenomena to become observable, the thermal energy $k_bT$ has to be much lower than the characteristic oscillation energy $\hbar\omega$. The flexural mode frequencies rarely exceed one GHz, and there are no such strongly nonlinear mechanical phenomena analogous to the Josephson tunneling. Hence, nonlinearity which helps to isolate quantum behavior becomes prominent only by the means of reducing the linear energy, with an accompanying reduction in frequency and stringent requirements for temperature. Also, the mechanical zero-point vibrations are orders of magnitude smaller than the typical length scales encountered in solid-state physics.

Macroscopic quantum tunneling (MQT) refers to quantum tunneling in a degree of freedom involving a macroscopic number of particles. This is how Josephson junctions were shown to portray quantum behavior more than 20 years ago, namely, by observing the phase to escape from the metastable minimum via MQT \cite{voss81,Jackel,martinis85}. Josephson junctions display strong nonlinearity at the zero-point energy scale, and the frequencies typically reside in the range of tens of GHz, and thus the quantum limit is encountered at relatively easily attainable temperatures below one Kelvin.

A possibility to induce nonlinearity into the mechanical potential energy was suggested in Refs.~\onlinecite{Wybourne,NoriPRB,NoriNJP}. In their setup, a mechanically induced longitudinal compression in a clamped beam would induce a double-well potential, and hence a possibility for macroscopic quantum tunneling of displacement of the beam buckling either left or right. While measurably high tunneling rates were predicted, the compression would have to be adjusted with extreme accuracy, and the crossover temperatures between thermal activation and MQT would fall in the microkelvin regime for beams longer than $L \sim 100$ nm.

In the present work, we discuss an alternative way of inducing nonlinearity into the  \emph{electro}mechanical potential at the scale of zero-point energy. We suggest to use a dc voltage to displace a conductive, clamped beam or membrane into a deflection close to the critical value where it gets pulled in to the electrode. Near this pull-in voltage, a metastable minimum appears in the electromechanical potential. The pull-in event can thus occur via a very fundamental physical process, that is, escape by quantum tunneling from a metastable state. We see that while the control of the deformation requires a high precision in the applied gate voltage, the crossover temperature becomes of the order of mK, and hence is within experimental reach.

In Sec.~\ref{meta} we first study the behavior of the flexural mode eigenfrequency when the beam is influenced by a dc electric field from a nearby voltage gate. In Sec.~\ref{pmqt} we calculate in detail the possibility of escaping from the metastable minimum via MQT, and consider in particular structures based on graphene, or carbon nanotubes. Starting from a simplified model, we derive analytical expressions for the quantities of interest in the limits of very thin or thick beam or sheet. Numerics are used to verify the results in a more general setting.

\section{Metastable minimum in the electromechanical potential}\label{meta}

\begin{figure}[ht]
\includegraphics[width=0.95\columnwidth]{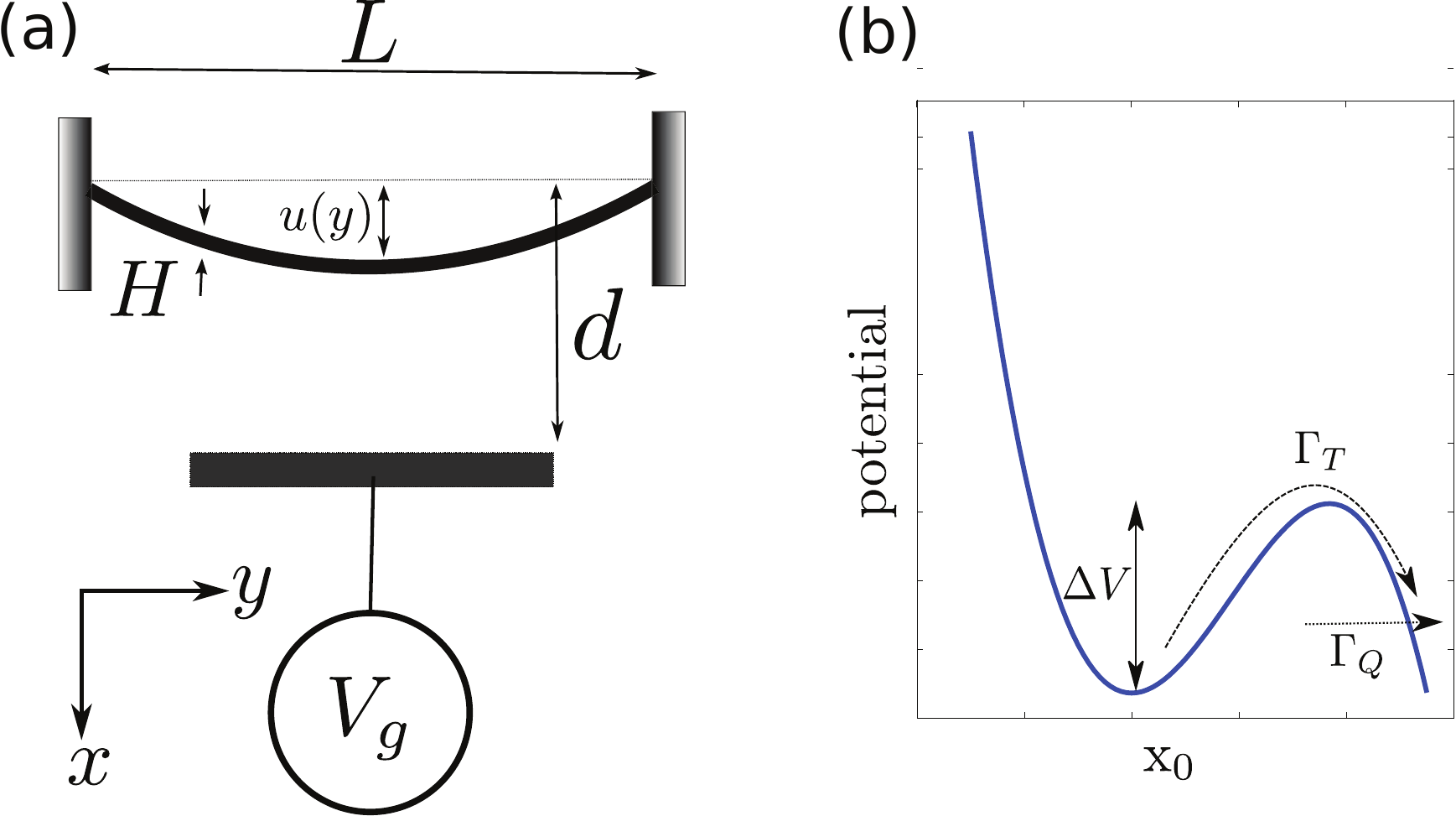}
\caption{(Color online)(a) Schematic picture of the studied nanoelectromechanical system. A beam or membrane is clamped from opposite ends, and its deformation is controlled by an electrostatic force created by a dc gate voltage $V_g$. (b) A metastable minimum in the potential energy near the pull-in point is introduced due to the nonlinearities in both electrical and mechanical energies. The system can escape from the minimum via thermal activation (rate $\Gamma_T$), or by macroscopic quantum tunneling (rate $\Gamma_Q$). } \label{NEMSqubit}
\end{figure}

We consider an electromechanical system modeled by a doubly clamped beam or sheet, having a length $L$, width $W$, thickness $H$, density $\rho$ and mass $m$. It is attached to the clamps by rigid boundary conditions. An electrostatic force created by a gate voltage $V_g$ situated at a distance $d$ from the beam  induces a deformation $u(y)$ to it, see Fig.~(\ref{NEMSqubit}). Note that we take the direction of the thickness $H$  along  the direction of deformation.

Let us write the potential energy of the system. For small displacements from the un-stretched position, the mechanical energy is quadratic with displacement. Here, however, we  consider large displacements nearly of the order $d$, and take into account the nonlinearity due to the elongation of the beam \cite{landaulf}. This affects the results substantially for thin membranes which stretch easily. We also include a built-in tension force $T_0$ which exists without the gate voltage and can be due to fabrication. In addition to the mechanical energy, the total potential energy has also an electrical contribution due to the voltage bias, given as $V_{\m{el}} = - \frac{1}{2} C_g[u(y)] V_g^2$, where $C_g$ is the capacitance between the beam and the gate. The sum of all these is
\begin{equation}\label{energy}
\begin{split}
\mathrm{E}_{\mathrm{total}}= & \frac{EI}{2}\int_0^L\left[\frac{d^2u(y)}{dy^2}\right]^2dy+\frac{T_0}{2}\int_0^L\left[\frac{du(y)}{dy}\right]^2dy\\
& +\frac{ES}{8L}\left(\int_0^L\left[\frac{du(y)}{dy}\right]^2dy\right)^2+V_{\m{el}},
\end{split}
\end{equation}
where $E$ is the Young modulus, $I$ the bending moment, $S$ the cross section of the beam. We now write the deformation of the beam as $u(y)=\mathrm{x}_0\chi(y)$, with $\mathrm{x}_0$ the amplitude and $\chi(y)$ the mode shape of the deformation. Henceforth, we consider the lowest flexural eigenmode, whose resonant frequency without the gate voltage is  $\omega_0^2=\frac{EI\mu}{m L^3}$, and the mode shape is given by a combination of trigonometric and hyberbolic functions \cite{clelandbook} (see Eq.~\eqref{frequncescaled} for definition of $\mu$).
Scaling the amplitude of the deformation with the beam-gate distance, $\xto=\mathrm{x}_0/d$, the energy becomes

\begin{equation}\label{j}
\begin{split}
\mathrm{E}_{\mathrm{total}}=& \frac{1}{2}m \omega_0^2d^2x_0^2+\frac{1}{2}m \omega_T^2 d^2 x_0^2\\
&+\frac{1}{2}m \omega_s^2 d^2x_0^4-\frac{V_g^2}{2}\int_0^LC_g[u(y)]dy,
\end{split}
\end{equation}
with $\frac{1}{2}m\omega_0^2 d^2=\frac{EI\mu}{2 L^3}d^2$ the bending energy of the beam displaced by $d$, $\frac{1}{2} m \omega_s^2 d^2=\frac{ESd^4\nu^2}{8L^3}$  the stress energy of the beam displaced by $d$  with respect to its equilibrium position and $\frac{1}{2}m \omega_T^2d^2=\frac{T_0}{2L}d^2\nu$ the stress energy related to initial tension. Here 
\begin{eqnarray}\label{frequncescaled}
\mu&=&\int_0^1\left[\frac{d^2\chi(y)}{dy^2}\right]^2dy,\\ 
\nu&=&\int_0^1\left[\frac{d\chi(y)}{dy}\right]^2dy,
\end{eqnarray}
depend on the shape of the deformation.
When using the approximation\cite{kozinsky} for the lowest flexural mode shape $\chi(y)=\sqrt{\frac{2}{3}}(1-\cos(2\pi y))$ we get $\mu=\frac{16 \pi^4}{3}$ and $\nu=\frac{4\pi^2}{3}$.

The pull-in phenomenon can be understood as follows. The gate capacitance $C_g(x)$, in general, increases towards an increasing deflection $x_0$ of the membrane. This causes an attractive force $F = -\partial_{x_o} \mathrm{E}_{\mathrm{total}}$ to appear between the membrane and the gate. The effective spring constant $ \partial^2_{x_0}\mathrm{E}_{\mathrm{total}}$ thus contains a positive (three first terms in Eq.~\eqref{j}) and a negative part (last term of Eq.~ \eqref{j}). Therefore it changes sign at a specific voltage $V_c$ dependent on the geometry, which corresponds to the membrane getting pulled in to contact with the gate. For example, in the simple plate capacitor model, the pull-in occurs at a static displacement $x_c = \frac{1}{3}$. In an energy picture, the minimum of the energy becomes metastable when increasing $V_g$ and disappears when $V_g=V_c$ as illustrated in
Fig.~2.
\begin{figure}[ht]
\includegraphics[width=0.8\columnwidth]{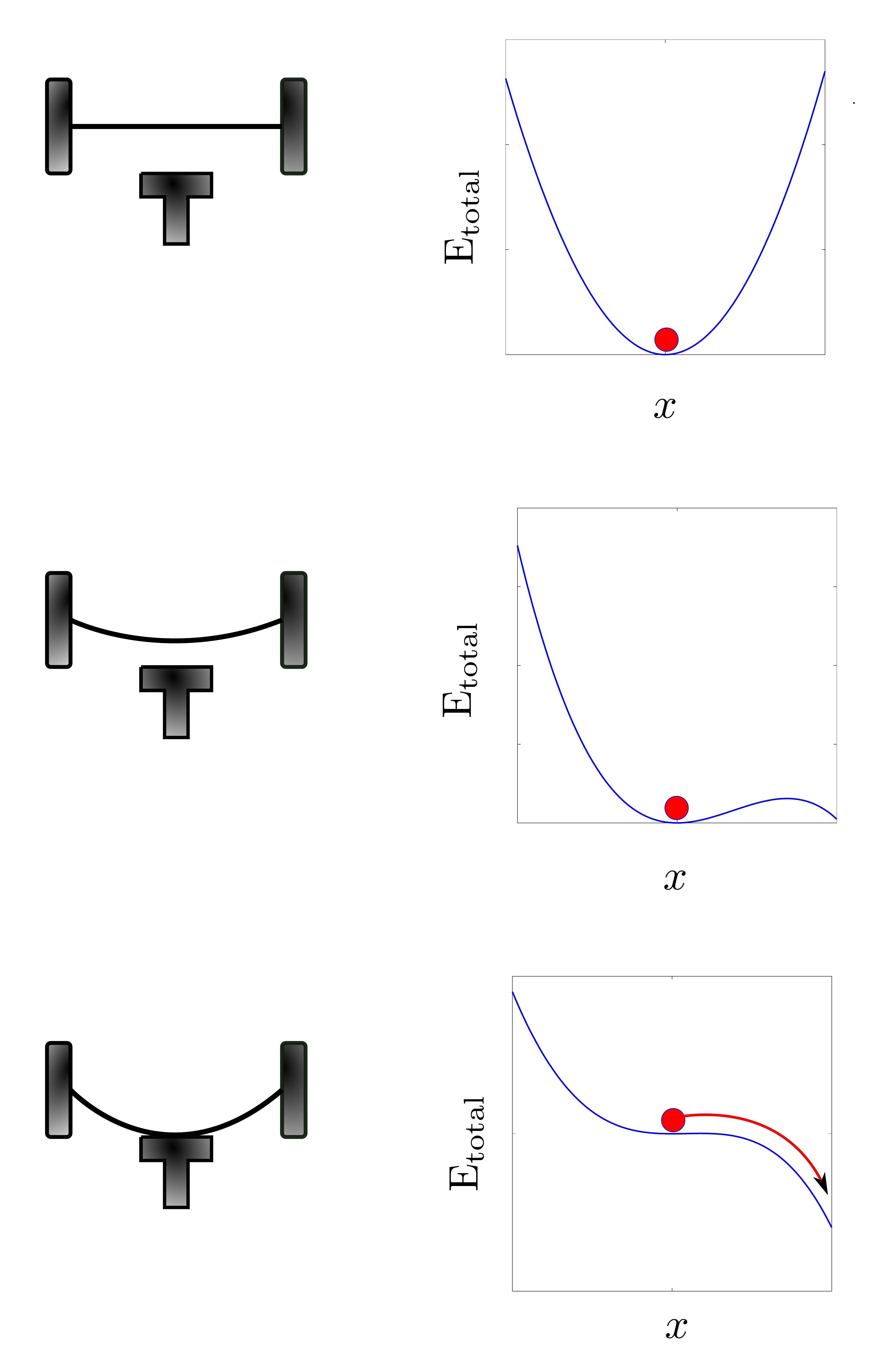}
\caption{(Color online) Schematic picture of the pull-in phenomenon in a suspended membrane or beam: when increasing the dc voltage, the amplitude of the deformation increases until the beam pulls into contact with the gate at a certain critical voltage $V_c$.}
\end{figure}

In order to specify our analysis, we have to choose a model for the capacitance. Qualitatively, however, the physics of the formation of the metastable minimum remains model-independent. We consider two cases: (A) the parallel plate capacitor, which is that of a membrane suspended above a back gate, and (B) a conducting wire parallel to a plane. We find carbon-based realizations as the most promising, due to their purity, stiffness and low mass. The cases A and B correspond, respectively, to graphene \cite{grapheneNEMS,stormer} and a single-wall carbon nanotube \cite{cnt,cnt2}.

First of all, in order to obtain simple analytical estimates, we assume that the capacitance depends only on the amplitude of the deformation $\mathrm{x}_0$. This approximation is relaxed in the numerical analysis, which takes into account the true shape of the deformation (see Appendix). Apart from the thick parallel plate model (section \ref{sec:thickplate}), $\mathrm{x}_0$ needs to be computed numerically from the force balance equation $-\nabla \mathrm{E}_{\mathrm{total}} = 0$.

\subsection{Parallel plate model}

Using the parallel plate capacitance model $C_g=\frac{\epsilon_0 W}{d(1-x_0)}$, the energy becomes

\begin{equation}\label{nrjpp}
\begin{split}
\mathrm{E}_{\mathrm{total}}=& \frac{1}{2}m \omega_0^2d^2x_0^2+\frac{1}{2}m \omega_T^2d^2 x_0^2\\
&+\frac{1}{2}m \omega_s^2 d^2x_0^4-\frac{V_g^2}{2}\frac{\epsilon_0WL}{d}\frac{1}{1-x_0}.
\end{split}
\end{equation}
Expanding Eq.~(\ref{nrjpp}) up to the third order for small variations $x$ around the equilibrium amplitude $x_0 (V_g)$  we obtain
\begin{equation}\label{forN}
\mathrm{E}_{\mathrm{total}}= \frac{1}{2}m\omega_x^2 d^2 x^2+\alpha x^3
\end{equation}
with
\begin{eqnarray}
\omega_x^2&=&\omega_0^2+\omega_T^2 \nonumber \\
&+&6\omega_s^2 \xto^2-V_g^2\frac{\epsilon_0WL}{md^3}\frac{1}{(1-x_0)^3}\label{wx}\\\label{alpha}
\alpha &=&2m\omega_s^2 d^2\xto-V_g^2\frac{\epsilon_0WL}{2d}\frac{1}{(1-x_0)^4}.
\end{eqnarray}
An inspection of Eq.~(\ref{wx}) reveals how the frequency depends on the gate voltage. The third term on the rhs, the mechanical nonlinearity, tends to increase the frequency with an increasing voltage, whereas the fourth term, electrical nonlinearity, has the opposite effect.
In experiments with carbon nanotubes \cite{cnt,cnt2} or graphene \cite{grapheneNEMS,stormer}, the frequency has typically been observed to go up, but for microfabricated metallic resonators, the frequency has decreased \cite{truitt,cnems}. We  now study the behavior of the frequency when we get close to the pull-in for two different regimes. In the first regime, we  consider the case where the bending energy is large compared to the stress energy, i.e $\frac{\omega_0}{\omega_s}\gg1 \;\Leftrightarrow\; \frac{d}{H}=\beta\ll1$. This is the case when
the variation of the frequency coming from the induced deformation is small, e.g., when the beam is  close to the gate or when the beam is  thick. We then consider
the opposite regime where it is the first term of Eq.~\eqref{wx} which can be neglected, e.g., for a thin beam or a beam far from the gate ($\frac{d}{H}\gg1$).

\subsubsection{Thick membrane, $\beta=\frac{d}{H} \ll 1$} \label{sec:thickplate}

For a beam close to the gate, the electrical potential is dominant, and the third term on the rhs of Eq.~(\ref{wx}) can be ignored.
This also holds if the built-in tension is large, since a larger voltage is then needed in order to obtain a certain $\mathrm{x}_0$. Scaling the energy with the bending energy we define  $\varepsilon_{\mathrm{thick}} \equiv \mathrm{E}_{\mathrm{total}} /(\frac{1}{2}m \omega_0^2d^2)$:
\begin{equation}
\varepsilon_{\mathrm{thick}}=\left(\frac{\omega_x}{\omega_0}\right)^2x^2-\frac{V_g^2}{\bar{V}^2}\frac{1}{(1-x_0)^4}x^3,
\end{equation}
\begin{figure}[h]
\center
\includegraphics[width=0.9\columnwidth]{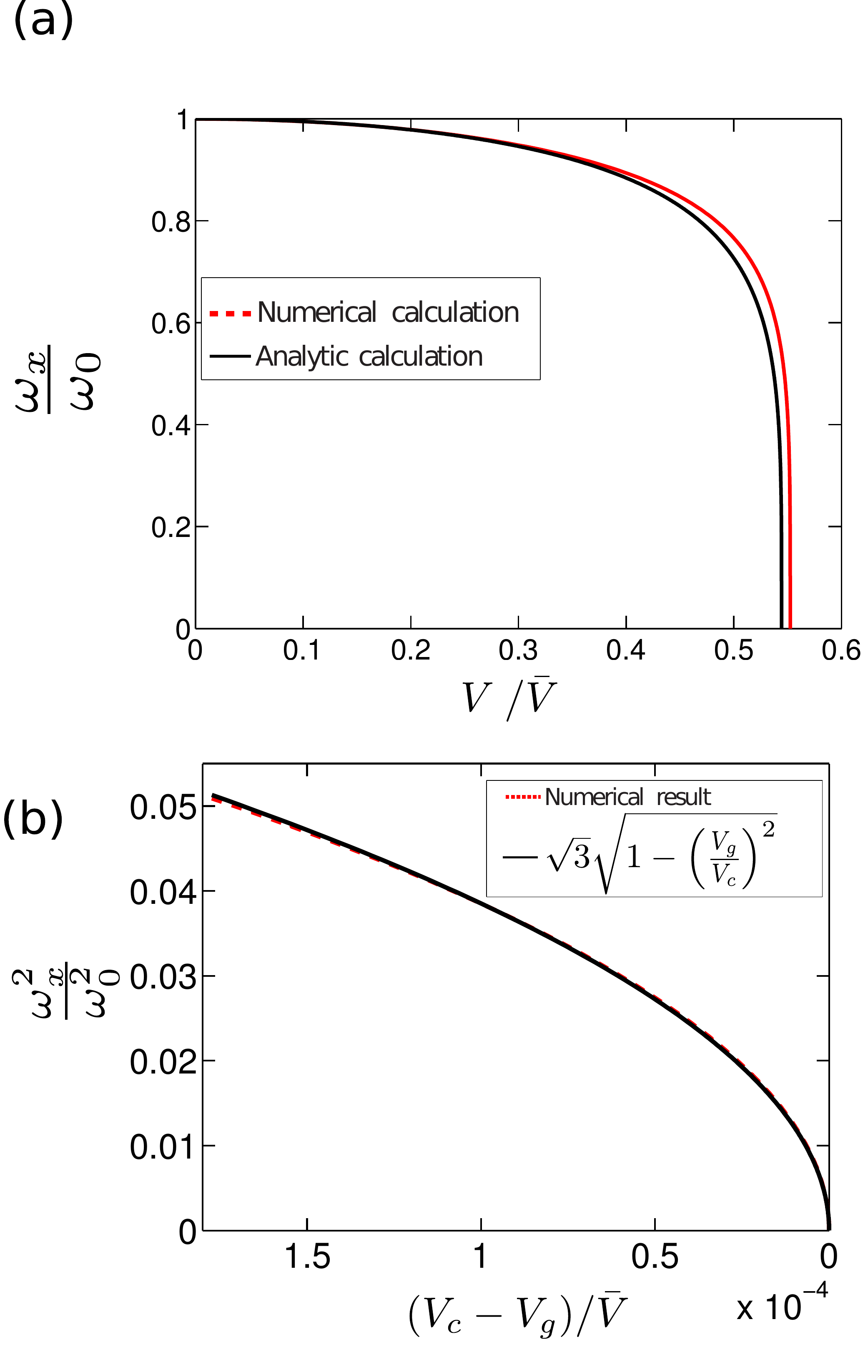}
\caption{(Color online) Behavior of the first flexural mode frequency with respect to the applied gate voltage scaled with $\bar{V}=\sqrt{\frac{ m \omega_0^2 d^3}{\epsilon_0 W L}}$ for a thick membrane having $T_0=0$, $L=W$ and $\beta=0.85$. (a) Black solid curve: analytical calculation supposing a constant mode shape, red dashed curve: full numerical calculation with the exact mode shape. (b) Behavior of the frequency close to the pull-in point.}\label{beta05}
\end{figure}%
with $\bar{V}^2=\frac{m\omega_0^2d^3}{\epsilon_0WL}$. The critical displacement at which pull-in point occurs is given by $x_c = 1/3$, \cite{kaajakari} and the critical voltage is
$V_c/\bar{V}=\sqrt{\frac{8}{27}}\left[1+\left(\frac{\omega_T}{\omega_0}\right)^2\right]$.
Using the plate-capacitor model, the position-dependent frequency and the third-order coefficient near the pull-in point are
\begin{eqnarray}
     \left(\frac{\omega_x}{\omega_0}\right)^2&=& \sqrt{3} \left[1+\left(\frac{\omega_T}{\omega_0}\right)^2\right]^2\sqrt{1-\left(\frac{{V_g}}{V_c}\right)^2}\label{eq:epotcoeff} \\
     \frac{\alpha }{m \omega_0^2 d^2}& = &  -\frac{3}{4} \left[1+\left(\frac{\omega_T}{\omega_0}\right)^2\right]^2\, . \label{eq:alphalin}
\end{eqnarray}
For example, a graphite resonator with $\rho=2$ g/cm$^3$, $L=W=0.5$ $\mu$m, $H=100$ nm, $d=10$ nm and $E=30$ GPa, would have $V_c=25.9$ V and $\omega_0/2\pi=1.6$ GHz. A numerical calculation of the behavior of the frequency with respect to the gate voltage is shown in Fig.~3. In this figure and Figs.~4-7, $d$ has been chosen  so that the accumulated strain in the pull-in position is at most half of the tensile strength.\cite{lee} 

\subsubsection{Thin membrane, $\beta=\frac{d}{H} \gg 1$}

The electrical potential term in Eq.~(\ref{energy}) is insignificant at small deflections, if the beam is thin or far from the gate. The beam then becomes
stiffer and the frequency goes up with the gate voltage as shown in Fig~4.
\begin{figure}[h]
\center
\includegraphics[width=0.9\columnwidth]{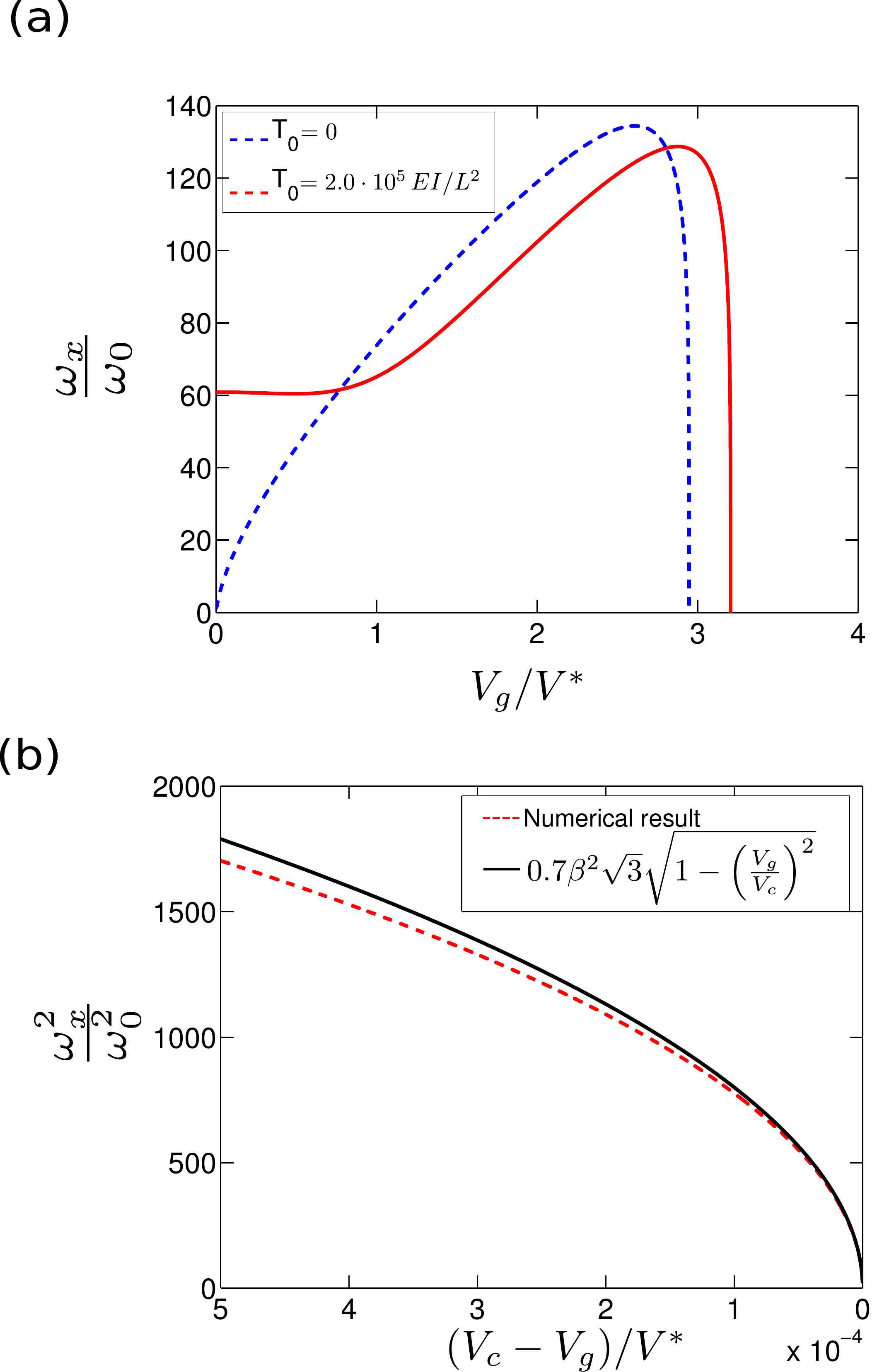}
\caption{(Color online) As Fig.~\ref{beta05}, but for a thin membrane with $(V^*)^2=\frac{m\omega_s^2d^3}{\epsilon_0WL}$, $L=W$ and $\beta=283$. (a) Full numerical calculation using the exact mode shape, with two different built-in tensions. The blue dashed curve: the initial tension is $T_0=0$ while for the red solid curve it is $T_0=2\cdot10^{5}\frac{EI}{L^2}$. (b) Fit of the numerical result close to the pull-in point with the equation $ a \beta^2\sqrt{3} \left[1-(V_g/V_c)^2 \right]^m$. The fit gives $a=0.7$ and $m=0.5$ which confirm the power law in Eq.~\eqref{eq:epotcoeffthin}.}\label{beta166}
\end{figure}

At larger deflections, however, the electrical term becomes dominant and the metastable potential minimum
becomes shallow and eventually disappears. In this case we may neglect the small constants $\omega_0^2$ and $\omega_T^2$  in Eq.~(\ref{wx}). The pull-in point is obtained when
the electrical and mechanical terms in Eq.~(\ref{wx}) cancel one another. Scaling the energy with the stress energy we define  $\varepsilon_{\mathrm{thin}}\equiv\mathrm{E}_{\mathrm{total}}/(\frac{1}{2} m\omega_s^2d^2)$:
\begin{equation}
\begin{split}
  \varepsilon_{\mathrm{thin}}=&\frac{1}{2}\left[12 x_0^2-2\left(\frac{V_g}{V^*}\right)^2\frac{1}{(1-x_0)^3}\right]x^2\\
&+\left[4x_0-\left(\frac{V_g}{V^*}\right)^2\frac{1}{(1-x_0)^4}\right]x^3,
\end{split}
\end{equation}

with $(V^*)^2=\frac{m\omega_s^2d^3}{\epsilon_0WL}$. We obtain the critical displacement $x_c = 3/5$, and the critical voltage\cite{numeric} $V_c/V^*=\frac{12}{25}\sqrt{\frac{3 }{5}}$. Near the pull-in point, we obtain from Eqs.~(\ref{wx},\ref{alpha})
\begin{equation}\label{eq:epotcoeffthin}
\begin{split}
     & \left(\frac{\omega_x}{\omega_0}\right)^2=\frac{ 54 \sqrt{\frac{3}{5}}}{5}\beta^2 \frac{\nu^2}{\mu}\sqrt{1-\left( \frac{V_g}{V_c} \right)^2}\\
    & \frac{\alpha }{m \omega_0^2 d^2} =-\frac{9}{2}\beta^2\frac{\nu^2}{\mu} .
\end{split}
\end{equation}
Contrary to a thick membrane, Eq.~(\ref{eq:alphalin}), the nonlinearity $\alpha$ depends on $\beta^2$. Therefore the nonlinearity is linked to the elongation of the beam.
For example, taking a single-layer graphene sheet with $\rho=2$ g/cm$^3$, $L=W=0.5$ $\mu$m, $H=0.3$ nm, $d=85$ nm and $E=1$ TPa, we get $V_c=17.8$ V and $\omega_0/2\pi=27.6$ MHz.

\subsection{Wire parallel to a plate}

The parallel plate capacitor model is a good approximation for membrane resonators, but another interesting case to consider is a suspended beam, such as a thin carbon nanotube. With the capacitance $C_g=\frac{2 \pi \epsilon_0 L}{\ln\left[2 d/H(1-x_0)\right]}$, $H$ being in this case the radius of the beam, the energy of the system is
\begin{equation}\label{nrjcnt}
  \mathrm{E}_{\mathrm{total}}= \frac{1}{2}md^2(\omega_0^2+\omega_T^2) x_0^2+\frac{1}{2}md^2\omega_s^2 x_0^4-V_g^2\frac{\pi \epsilon_0L}{\ln\left[2\beta \left(1-x_0\right)\right]}.
\end{equation}
As in the previous section we make an expansion
of Eq.~(\ref{nrjcnt}) up to the third order for small variation $x$ around the equilibrium amplitude $x_0 (V_g)$  and obtain
\begin{equation}
 \mathrm{E}_{\mathrm{total}}= \frac{1}{2}m\omega_x^2d^2 x^2+\alpha  x^3,
\end{equation}
with
\begin{eqnarray}
\omega_x^2&=&\omega_0^2+ \omega_T^2+6\omega_s^2\xto^2 \nonumber \\
&-&\frac{V_g^2}{md^2}\frac{\pi \epsilon_0L \{\ln [-2 (x_0-1) \beta ]+2\}}{(x_0-1)^2 \ln ^3[2 (1-x_0) \beta ]}\label{wxcnt}\\ \label{alphacnt}
\alpha &=&2m\omega_s^2 d ^2 x_0\\
&-&\frac{V_g^2\pi\epsilon_0  \left(\ln ^2(2 (1-x_0) \beta )+3 \ln (2 (1-x_0) \beta )+3\right)}{3 (x_0-1)^3 \ln ^4(2 (1-x_0) \beta )}\nonumber.
\end{eqnarray}

The behavior of the frequency is similar to the one found in the previous section. Below we only consider the case of a thin tube $\frac{d}{H}\gg1$ which best corresponds to carbon nanotubes \cite{Sapmaz}.

\subsubsection{Thin tube, $\beta=\frac{d}{H} \gg 1$}

As in the previous section, we may neglect the small constant $\omega_0^2$ term in Eq.~(\ref{wxcnt}). The pull-in point is obtained when
the electrical and mechanical terms in Eq.~(\ref{wxcnt}) cancel one another. Neglecting in the derivatives the $x_0$ dependence in the logarithms, we obtain the critical displacement $x_c =\frac{ 3 \ln{2 \beta}}{2 (1 + 2 \ln{2 \beta})}$  and the critical voltage
$(V_c/\tilde{V})^2= \frac{27 \beta^2  \ln{2 \beta}^5 (2+\ln{2 \beta})}{4  (1+\ln{4}+2 \ln{\beta})^4}$ with $\tilde{V}^2=\frac{m\omega_s^2 d^3}{2\pi\epsilon_0 L}$. Near the pull-in point, we obtain from Eqs.~(\ref{wxcnt},\ref{alphacnt}):
\begin{equation}\label{eq:epotcoeffthint}
\begin{split}
     &\left(\frac{\omega_x}{\omega_0}\right)^2=\frac{9  \beta^2 \nu^2 \ln{2 \beta}^2}{2 \mu (1 + \ln{4} + 2 \ln{\beta})}\left(1- \frac{V_g}{V_c}\right)\\
    & \frac{\alpha }{m \omega_0^2 d^2} = \frac{3 \beta ^2 \nu ^2 \ln (2 \beta ) }{8 \mu ^2}C,\\
\end{split}
\end{equation}
where
\begin{equation*}
 C=\frac{4 \mu  }{ \ln (\beta )} - \frac{81 \beta ^2 \nu ^2 \ln ^7(2 \beta )}{ 32 \ln^6 (\beta )}.
\end{equation*}

Similarly to the thin membrane, the nonlinearity is due to the elongation of the beam, but there is a difference in the behavior of the frequency close to the pull-in. Here, $\omega_x^2$ decreases linearly with the voltage, instead of being proportional to a square root of $V_g$ as in Eq.~\eqref{eq:epotcoeff}. Therefore, the pull-in for a beam is expected to occur in a ``smoother'' way than for a membrane. The results for the frequency are summarized in Fig~5. Although the qualitative behavior of the frequency is mostly captured by the analytical calculation, very close to
the pull-in we again capture the square root behavior (Fig.~5b). For example, with a single-wall carbon nanotube resonator with $\rho=2$ g/cm$^3$, $L=0.5$ $\mu$m, $H=1$ nm, $d=85$ nm and $E=1$ TPa, we have $V_c=6.2$ V and $\omega_0=23$ MHz.
\begin{figure}[h]
\center
\includegraphics[width=0.9\columnwidth]{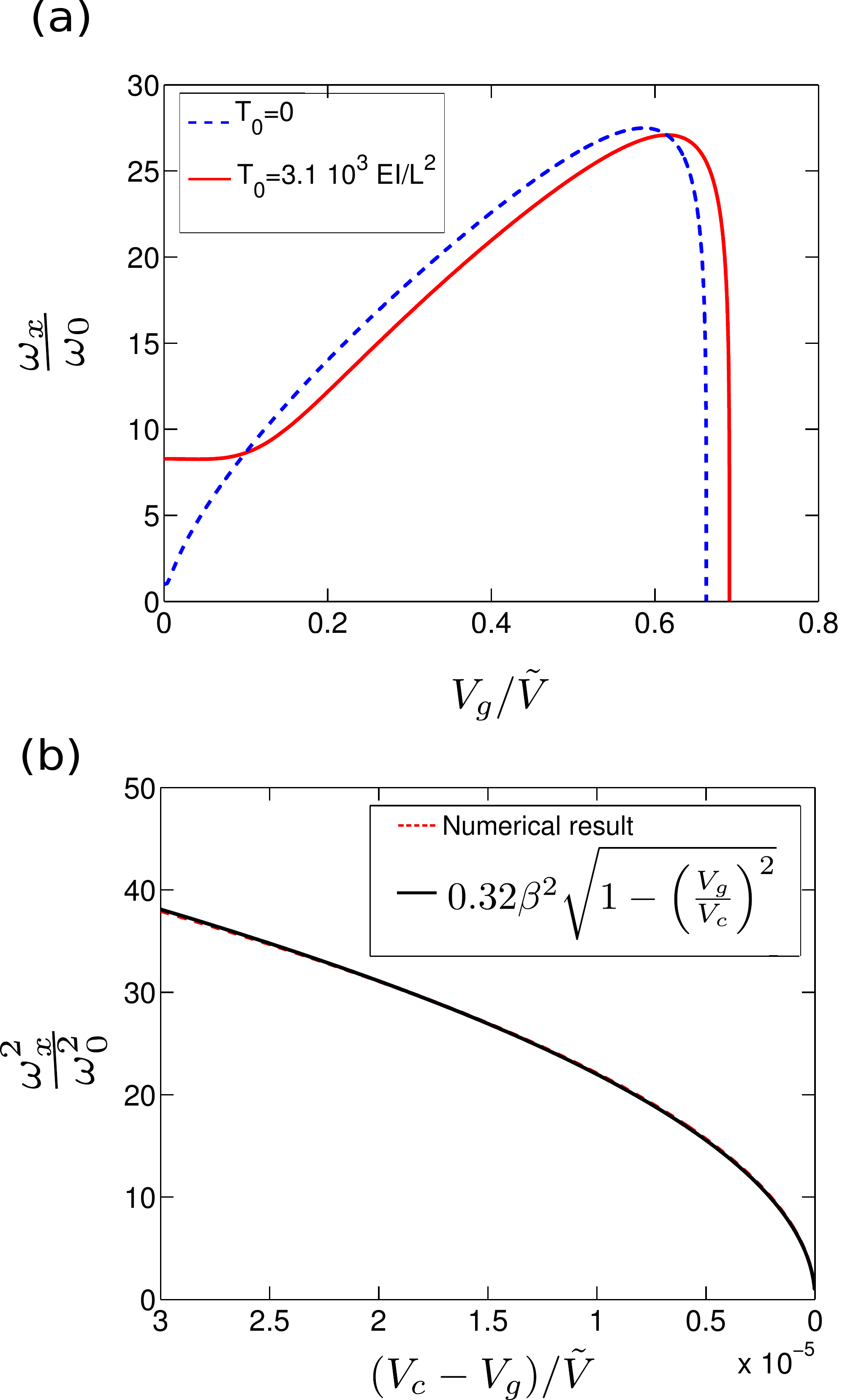}
\caption{(Color online) The fundamental flexural mode frequency with respect to the applied gate voltage with $\tilde{V}^2=\frac{m\omega_s^2 d^3}{\pi\epsilon_0 L}$ for a single-wall carbon nanotube with
$\beta=85$. (a) The blue dashed curve is without an initial tension ($T_0=0$), while the red solid curve is with $T_0=3.1 \cdot10^{3}\frac{EI}{L^2}$. (b) Fit of the numerical result with the equation $ a \beta^2(1-(V_g/V_c)^2)^m$ indicates a power law with $a=0.32$ and $m=0.5$.}\label{wxTube}
\end{figure}

\section{Prospects of observing MQT}\label{pmqt}

\begin{figure}[h]
\center
\includegraphics[width=0.8\columnwidth]{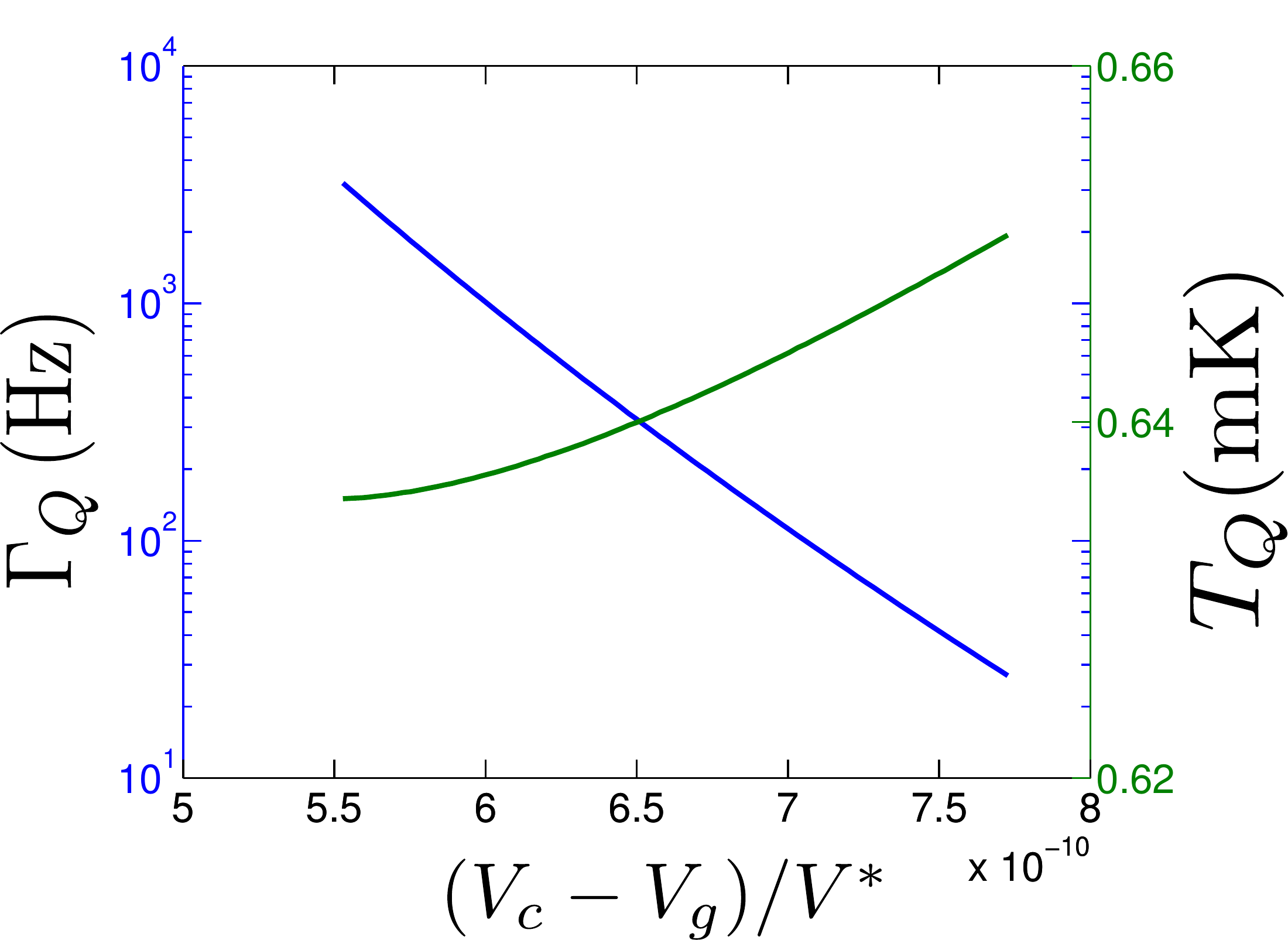}
\caption{(Color online) Macroscopic quantum tunneling (MQT) in a thin membrane, $\beta \gg 1$. Right hand axis in green: the quantum-classical crossover temperature,  plotted as a function of departure of the dc gate voltage from the critical voltage with $V^{*^2}=\frac{m \omega_s^2 d^3}{\epsilon_0WL^4}$. Left hand axis in blue: the rate of MQT of the displacement out of the metastable minimum. The plots are for a representative sample consisting of a single-layer graphene sheet with $H = 0.3$ nm, $L=5 \, \mu$m, $W=2 \, \mu$m, $E=1$ TPa, $\rho = 2$ g/cm$^3$,  and $d=85$ nm. The lowest number of states is $N \sim 2$.} \label{MQTrateGraph}
\end{figure}

\begin{figure}[h]
\center
\includegraphics[width=0.8\columnwidth]{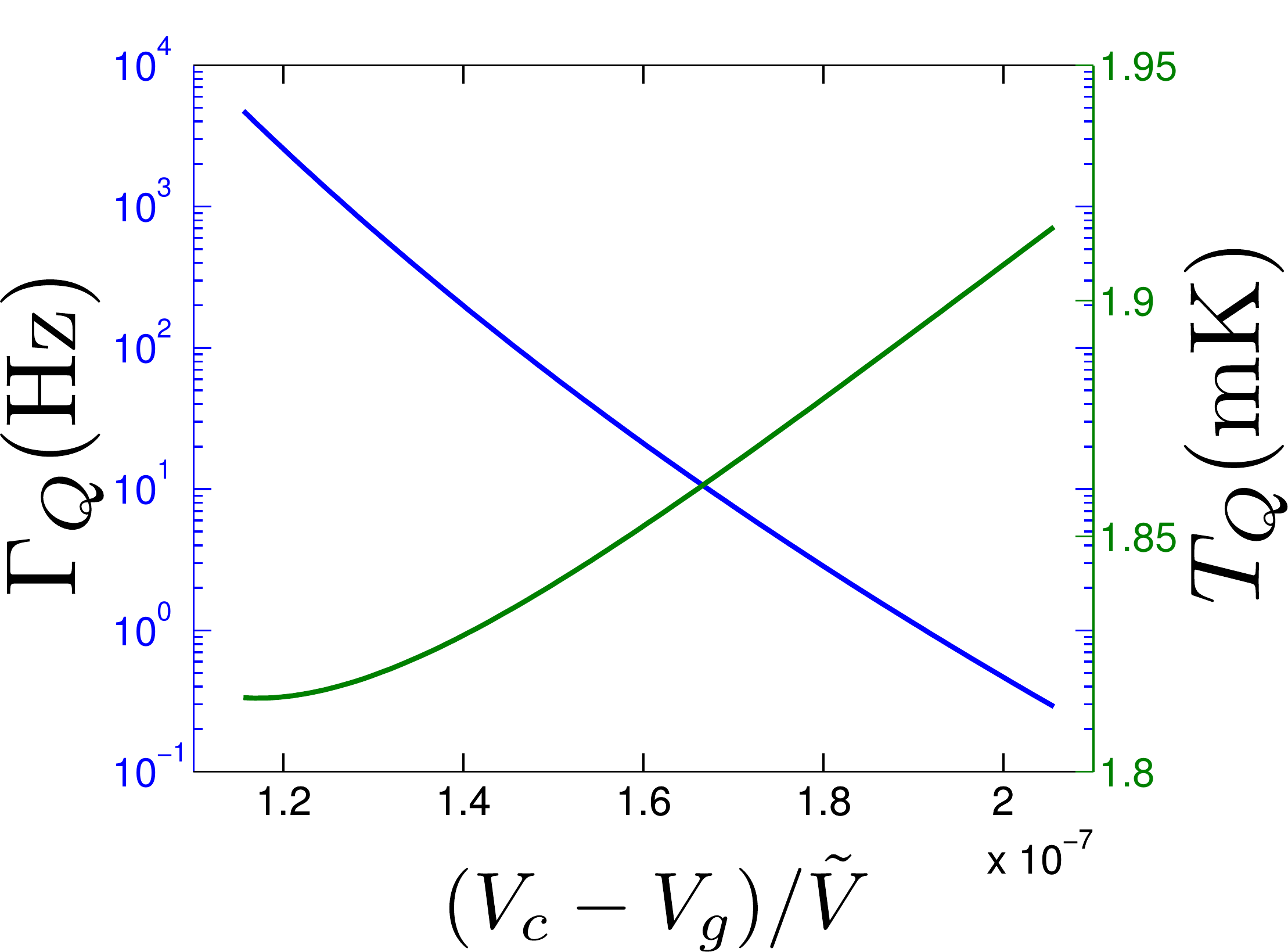}
\caption{(Color online) As Fig.~\ref{MQTrateGraph}, but for a carbon nanotube, with $H = 1$ nm, $L=0.5 \, \mu$m, $E=10^3$ GPa, $\rho = 2$ g/cm$^3$,  and $d=85$ nm. Here $\tilde{V}^2=\frac{m\omega_s^2 d^3}{2\pi\epsilon_0 L}$.} \label{MQTrateCNT}
\end{figure}

At a temperature $T$, the rate of thermally activated escape from the metastable minimum in Fig.~3 is given by the usual Arrhenius law
$\Gamma_T = \omega_x /(2\pi) \exp \LL( -\Delta V /k_B T \RR)$ where $\omega_x$ is the frequency of the resonator. The tunneling rate from the ground state is given by\cite{weiss}
$\Gamma_Q = \omega_Q  \exp\LL( -\frac{36}{5} \frac{N}{2\pi} \RR)$, where $N =  \frac{\Delta V}{\hbar \omega_x}$ is the number of states in the potential with height $\Delta V$, and $\omega_Q=6\omega_x \sqrt{6N/\pi}$. The crossover temperature $T_Q$ below which quantum tunneling from the ground state dominates over thermal escape is given by equating the Arrhenius law with the quantum tunneling rate,
%
\begin{equation}\label{eq:TQ}
T_Q = \frac{5}{36}\frac{\hbar \omega_x}{k_b}\frac{1}{1-\frac{5}{36}\frac{1}{N}\ln\left(12 \sqrt{\frac{6}{\pi}N}\right)}.
\end{equation}
In addition to giving the crossover between the ground-state tunneling and thermal activation, this is also a good criterion for the system
to be in the quantum limit in the sense that thermal population becomes negligible. Another important quantity is  the number $N$ of bound states in the metastable minimum. From Eq.~(\ref{forN}) we obtain
\begin{equation}\label{eq:N}
    N \simeq \frac{\Delta V}{\hbar \omega_x} =\frac{1}{54} \frac{d^6 m^3\omega_x^{5}}{\hbar\alpha^2}.
\end{equation}
In order to maximize the quantum tunneling rate we need to have $N\sim1$.
For a membrane in the case where $\beta\ll1$ the frequency at which the number of states
is close to 1 satisfies
\begin{equation}
\left(\frac{\omega_x}{\omega_0}\right)^{5}=\frac{3^5}{8}\frac{1}{N_0},
\end{equation}
while  when $\beta\gg1$ it is
\begin{equation}
\left(\frac{\omega_x}{\omega_0}\right)^5\approx 1100\frac{\nu^4}{\mu^2}\frac{ \beta^4}{N_0 }\label{frequencethin},
\end{equation}
with $N_0=m \omega_0 d^2/\hbar$. From these equations we can see that the mass of the resonator has to be the lowest possible in order to maximize the frequency, which would then lead to the highest crossover temperature, and hence the best experimental prospect of observing MQT. Also Eq.~(\ref{frequencethin}) shows that a system having a large $\beta$ is favorable since it yields a higher frequency when the number of states is close to 1. The most suitable materials from both of these points of view, having a low mass and possibility for a high $\frac{d}{H}$ ratio, are graphene and carbon nanotubes. In Figs.~\ref{MQTrateGraph} and ~\ref{MQTrateCNT} we plot a full numerical calculation of the crossover temperature and the MQT rate for graphene and CNT resonators, which show the conflicting requirements of attaining a high frequency and simultaneously having a measurable tunneling rate.

MQT could be verified by measuring the reduction of the escape rate as a function of temperature, and observing its saturation at the crossover
temperatures predicted above. An escape event would easily be detected as a large change of capacitance once the membrane gets pulled in, for instance, by using the dispersive methods \cite{lehnert,cnems,Schwab10,FIB,Teufel11} which do not otherwise excessively disturb the system. In order to repeat the experiment, the gate voltage would be reset to zero, and another gate on the opposite side could be used to pull the membrane from the Van der Waals attraction.   

A possible experimental verification is challenged by the fact that reaching a small number of states requires a high precision on the applied gate voltage, and a small change in the applied gate voltage leads to a large variation in the frequency in this regime. The wire parallel to a plate model shows that the variation close to the pull-in point is smoother than for the parallel plate model, and in this respect,  carbon nanotubes seem to be  good candidates for observing MQT.

\section{Conclusions}

Macroscopic quantum tunneling (MQT) is a fascinating topic which allows one to investigate the quantum-classical transition regime, where frequencies of collective degrees of freedom become comparable to temperature. We have introduced a model system of a mechanical degree of freedom trapped into a metastable state formed by a conductive beam or membrane suspended on a voltage biased back gate. Via MQT, the fictitious particle can escape from the metastable state and thereby be pulled into contact with the gate. The setup might serve as a means of observing mechanical MQT, in a fashion analogous to how quantum behavior in Josephson systems was first observed \cite{voss81,Jackel,martinis85}.

Here we discuss the possibility of observing MQT in nanomechanical resonators for two capacitance models, those of a parallel plate capacitor, and the wire parallel to a plate, corresponding, for example, to suspended graphene or carbon nanotube. One should have the number of states in the metastable minimum close to one in order to obtain a sufficient quantum tunneling rate, and at the same time, maintain a high frequency to maximize the crossover temperature. The highest crossover temperatures are obtained in a system with low mass, low density, high Young's modulus, and a high ratio $d/H$ which allows to increase the frequency by increased strain. These criteria point towards carbon-based realizations.

We conclude that while the predicted crossover temperatures in the mK-range are several orders of magnitude higher than for the buckled beam studied previously, they are still experimentally demanding, and barely within reach of standard dilution refrigerator techniques. However, one may use electrical cooling techniques where the nanoresonator is coupled to higher-frequency electrical resonator in order to cool the lowest mode in question down to temperatures much lower than the environment \cite{Gigan,Schwab10}. We thus foresee the experimental verification challenging, but possible in the future.

\begin{acknowledgments}
This work was supported by the Academy of Finland, and jointly by the NSF under DMR-0908634 (Materials World Network), and by the European Research Council (grants No. 240362-Heattronics, 240387-NEMSQED and EU-FP7-NMP-246026).
\end{acknowledgments}

\section*{Appendix}\label{appendix}
For the numerical calculations, we assume that there is a small deviation $\delta u(y,t)=u_1(y) e ^{i \omega t}$ from the static deformation $u_0(y)$ such that the total deformation can be written in the form
\begin{equation}
u(y,t)=u_0(y)+\delta u(y,t).
\end{equation}
Introducing these expressions in the Euler-Bernoulli
equation, neglecting the terms which are  $O(\omega^2)$ and taking the parallel plate capacitance model leads to the  equation for DC deflection $u_0$
\begin{multline}\label{u0}
\frac{\partial^{4}u_0(y)}{\partial y^4}-\left(\alpha+6 \beta^2\int_{0}^{1}u_0'(z)^{2}dz\right)\frac{\partial^{2}u_0(y)}{\partial y^2}\\
-\tilde{V}^2 \frac{1}{(1-u_0)^2} = 0.
\end{multline}
The AC part $u_1(y,t)$ satisfies an eigenvalue equation for $u_1$
\begin{multline}\label{u_1}
\frac{\partial^{4}u_1(y)}{\partial y^4}-\left(\alpha+6 \beta^2\int_{0}^{1}u_0'(z)^{2}dz\right)\frac{\partial^{2}u_1(y)}{\partial y^2}\\
-2\tilde{V}^2\frac{u_1}{(1-u_0)^3}
-12\beta^2 \vec{u_0''}(y)\otimes \vec{u_0'}(y)\frac{\partial u_1(y)}{\partial y}= \frac{\omega^2}{\omega_0^2}\,u_1.
\end {multline}
Here,
\begin{eqnarray}
\alpha&=&\frac{L^2}{EI}T_0\\
\beta&=&\frac{d}{H}\\
\tilde{V}^2&=&V_g ^2 6\left(\frac{d}{H}\right)^3\frac{\epsilon_0 L^4}{d^6}\\
\omega_0^2&=&\frac{EI}{mL^3}.
\end{eqnarray}

Writing $u_0=\sum_i^n a_i \chi_i(y)$ with $\chi_i(y)$ the $i^{th}$ flexural eigenmode, \cite{clelandbook} we solve  \eqref{u0}
using the Galerkin method,  rewriting \eqref{u_1} in the eigenmode space ($\chi_i$-space) and use the solution found for $u_0$ to compute  $\omega$ and $u_1$. This first step allows us to find the behavior of the frequency with respect to the applied gate voltage. This is  shown
in Figs.~\ref{beta05} and \ref{beta166} in the thick and thin membrane regime limits,  for $\beta<1$ and $\beta>1$, respectively.
We then use $u_0(y)$ and $u_1(y,t)$ and use Eq.~\eqref{energy} to plot $\varepsilon(u_0+ \mathrm{x}_0 u_1)$ for a particular gate voltage, $\mathrm{x}_0$ being the amplitude of the AC deformation. This step allows
us to compute the height of the metastable potential $\Delta V$. From these results it is possible to compute the number of states and thus the quantum tunneling rate as discussed in Sec. \ref{pmqt}.

\newpage

\begin{thebibliography}{99}
\bibitem{LaHaye04}M. D. LaHaye, O. Buu, B. Camarota, and K. C. Schwab, Science \textbf{304}, 74 (2004).
\bibitem{SchwabQ}K. C. Schwab and M. L. Roukes, Phys. Today \textbf{58}, 36 (2005).
\bibitem{Kippenberg08} T. J. Kippenberg and K. J. Vahala, Science \textbf{321}, 1172 (2008).
\bibitem{ClelandMartinis} A. D. O'Connell \emph{et al.}, Nature \textbf{464}, 697 (2010).
\bibitem{voss81} R. F. Voss and R. A. Webb, Phys. Rev. Lett. \textbf{47}, 265 (1981).
\bibitem{Jackel} L. D. Jackel \emph{et al.}, Phys. Rev. Lett. \textbf{47}, 697 (1981).
\bibitem{martinis85} J. M. Martinis, M. H. Devoret, and J. Clarke, Phys. Rev. Lett. \textbf{55}, 1543 (1985).
\bibitem{Wybourne} S. M. Carr, W. E. Lawrence, and M. N. Wybourne, Phys. Rev. B \textbf{64}, 220101 (2001).
\bibitem{NoriPRB} S. Savel'ev, A. L. Rakhmanov, X. Hu, A. Kasumov, and F. Nori, Phys. Rev. B \textbf{75}, 165417 (2007).
\bibitem{NoriNJP} S. Savel'ev, X. Hu, and F. Nori, New J. Phys. \textbf{8}, 105 (2006).
\bibitem{landaulf} L. D. Landau and E. M. Lifshits, \emph{Theory of Elasticity} (Pergamon, Oxford, 1986).
\bibitem{clelandbook} A. Cleland, \emph{Foundations of Nanomechanics} (Springer, New York, 2003).
\bibitem{kozinsky} I. Kozinsky, H. W. Ch. Postma, I. Bargatin, and M. L. Roukes, Appl. Phys. Lett. \textbf{88}, 253101 (2006).
\bibitem{grapheneNEMS} J. S. Bunch, \emph{et al. }, Science \textbf{315}, 490 (2007).
\bibitem{stormer} C. Chen \emph{et al.}, Nature Nanotechnology \textbf{4}, 861 (2009).
\bibitem{cnt} V. Sazonova, Y. Yaish, H. Ustunel, D. Roundy, T. A. Arias, and P. L. McEuen, Nature \textbf{431}, 284 (2004).
\bibitem{cnt2} B. Witkamp, M. Poot, and H. S. J. van der Zant, Nano Lett. \textbf{6}, 12 (2006).
\bibitem{truitt} P. A. Truitt, J. B. Hertzberg, C. C. Huang, K. L. Ekinci, and K. C. Schwab, Nano Lett. \textbf{7}, 120 (2007).
\bibitem{cnems} M. A. Sillanp\"a\"a, J. Sarkar, J. Sulkko, J. Muhonen, and P. J. Hakonen, Appl. Phys. Lett. \textbf{95}, 011909 (2009).
\bibitem{kaajakari} see, e.g., V. Kaajakari, \emph{Practical MEMS} (Small Gear Publishing, 2009).
\bibitem{lee} Changgu Lee, Xiaoding Wei, Jeffrey W. Kysar, James Hone, Science, \textbf{18}, 321 (2008).
\bibitem{numeric} Note that this analytical approximation overestimates the critical voltage and the critical displacement. This is  due to the neglect of the contribution of higher order modes in the presence of large displacements, captured by our numerical model.
\bibitem{Sapmaz}S. Sapmaz and Ya. M. Blanter and L. Gurevich and H. S. J. van der Zant, Phys. Rev. B. \textbf{67}, 235414 (2003).
\bibitem{weiss} U. Weiss, \emph{Quantum Dissipative Systems} (World Scientific, Singapore, 1999).
\bibitem{lehnert} C. A. Regal, J. D. Teufel, and K. W. Lehnert, Nature Physics \textbf{4}, 555 (2008).
\bibitem{Schwab10} T. Rocheleau, T. Ndukum, C. Macklin, J. B. Hertzberg, A. A. Clerk, and K. C. Schwab, Nature \textbf{463}, 72 (2009).
\bibitem{FIB} J. Sulkko, M. A. Sillanp\"a\"a, P. H\"akkinen, L. Lechner, M. Helle, A. Fefferman, J. Parpia, and P. J. Hakonen, Nano Letters \textbf{10}, 4884 (2010).
\bibitem{Teufel11} J. D. Teufel, Dale Li, M. S. Allman, K. Cicak, A. J. Sirois, J. D. Whittaker, and R. W. Simmonds, Nature \textbf{471}, 204 (2011).
\bibitem{Gigan} S. Gigan \emph{et al.}, Nature \textbf{444}, 67 (2006).











\end{thebibliography}

\end{document}